\documentclass[aps,prd,twocolumn,superscriptaddress,nofootinbib]{revtex4-2}

\usepackage{amsmath,amssymb}
\usepackage{graphicx}
\usepackage[dvipsnames]{xcolor}

\IfFileExists{orcidlink.sty}{\usepackage{orcidlink}}{}
\IfFileExists{ulem.sty}{\usepackage[normalem]{ulem}}{}
\providecommand{\orcidlink}[1]{}

\IfFileExists{caption.sty}{\usepackage[font=small, labelfont=bf,
  justification=raggedright, singlelinecheck=false]{caption}}{}
\IfFileExists{booktabs.sty}{\usepackage{booktabs}}{}
\IfFileExists{enumitem.sty}{\usepackage{enumitem}}{}
\IfFileExists{subcaption.sty}{\usepackage{subcaption}}{}

\usepackage{hyperref}

\newcommand{\dd}{\mathrm{d}}

\newcommand{\eV}{\ensuremath{\,\mathrm{eV}}}
\newcommand{\keV}{\ensuremath{\,\mathrm{keV}}}
\newcommand{\MeV}{\ensuremath{\,\mathrm{MeV}}}

\newcommand{\AU}{\ensuremath{\,\mathrm{AU}}}

\newcommand{\change}[1]{{#1}}

\definecolor{apsblue}{RGB}{0,0,128}
\hypersetup{
  colorlinks = true,
  allcolors  = apsblue
}

\begin{document}

\title{Enhanced Stellar Production of Weakly Interacting Slim Particles from Non-Thermal Nuclear Cascades}

\author{V\'ictor Fonoll~\orcidlink{0009-0003-1740-4474}}
\email{vfonoll@unizar.es}
\affiliation{CAPA \& Departamento de F\'isica Te\'orica, Universidad de Zaragoza, C. Pedro Cerbuna 12, 50009 Zaragoza, Spain}

\author{Maurizio Giannotti~\orcidlink{0000-0001-9823-6262}}
\email{mgiannotti@unizar.es}
\affiliation{CAPA \& Departamento de F\'isica Te\'orica, Universidad de Zaragoza, C. Pedro Cerbuna 12, 50009 Zaragoza, Spain}

\author{Giuseppe Lucente~\orcidlink{0000-0002-3266-3154}}
\email{lucenteg@slac.stanford.edu}
\affiliation{SLAC National Accelerator Laboratory, Stanford University, Menlo Park, CA 94025}

\begin{abstract}


Weakly interacting slim particles (WISPs) can be produced in stars through the conversion of non-thermal photons generated in nuclear reactions. Previous studies have generally treated these sources only at the level of their primary injection lines. We show that this picture is incomplete: repeated Compton scatterings redistribute the injected photons into a broad low-energy spectrum, while associated positrons can thermalize and annihilate into a 511~keV line. Together, these effects define a generic non-thermal photon reservoir and thus a broadly applicable source term for any photon-coupled WISP. We develop a general framework for this mechanism and illustrate its impact with the example of dark-photon production in the solar pp chain. Our results show that non-thermal stellar WISP production can be substantially underestimated if Compton reprocessing and positron annihilation are neglected. \change{We further identify the energy window above $\simeq90$~keV, where this reprocessed reservoir becomes the dominant source of solar WISPs, and we characterize the spectral signature it predicts there.}
\end{abstract}
\maketitle


\section{Introduction}

An extensive worldwide effort is under way to explore the low-energy
frontier of particle physics, where new light and feebly interacting
particles may reside.
Weakly interacting slim particles
(WISPs)~\cite{Jaeckel:2010ni,Ringwald:2012hr}---including axion-like
particles, dark photons, and other hidden-sector states---have emerged
as leading candidates to explain dark matter, as mediators of new
forces, or simply as generic low-energy relics of ultraviolet
completions of the Standard Model.
The theoretical motivations and the experimental landscape have been
thoroughly studied in a number of dedicated papers and monographs, and surveyed in recent community white
papers~\cite{Antel:2023hkf,Agrawal:2021dbo,Arza:2026rsl}, which
document a broad and rapidly growing program of searches spanning
laboratory experiments, cosmological observations, and astrophysical
probes.

Astrophysical environments provide powerful laboratories to probe such
light and weakly interacting particles.
WISPs can be produced in stellar interiors through their couplings to
photons, electrons, or nuclei~\cite{Raffelt:1996wa,Arza:2026rsl,Carenza:2024ehj}.
Most previous studies have focused on \emph{thermal} production
mechanisms determined by the equilibrium properties of the plasma,
including plasmon conversion, bremsstrahlung, and Compton-like
processes in stellar interiors.
However, stars also host \emph{non-thermal} sources associated with
nuclear reactions, which inject photons and charged particles with
energies far above the local thermal scale $T_\odot$.
A prominent example is the radiative deuterium capture in the solar pp chain,
\begin{equation}
  p + d \to {}^{3}\mathrm{He} + \gamma\,,
  \label{eq:pd_capture_intro}
\end{equation}
which produces a monoenergetic $5.49\;\mathrm{MeV}$ photon that can convert into a dark photon or an axion-like particle~\cite{CAST:2009klq,Borexino:2012guz,Bhusal:2020bvx,DEramo:2023buu}.
To date, analyses of such channels have retained only the primary nuclear photon line, 
neglecting the electromagnetic processing that follows the initial injection.
In a stellar plasma, however, direct escape through WISP conversion is unlikely. 
Instead, the injected photon typically undergoes successive Compton scatterings, each carrying a finite probability of WISP conversion,
generating a lower-energy continuum---a Compton cascade---that significantly extends the WISP spectrum.
\begin{figure}
    \centering
    \includegraphics[width=1\linewidth]{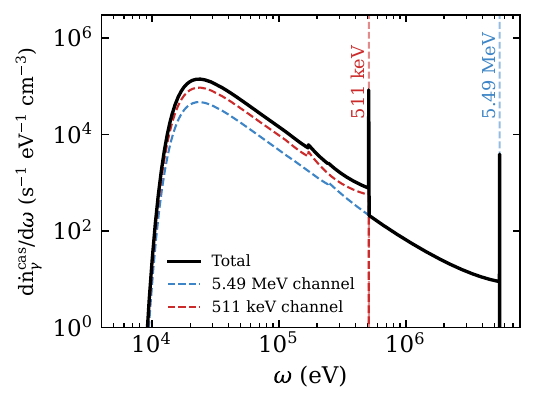}
\caption{
Local non-thermal differential photon source, $\dd\dot n_\gamma^{\rm cas}/\dd\omega$, defined in Eq.~\eqref{eq:X_flux_total}, 
generated by the Compton
cascade of the two pp-chain injection lines and evaluated at a representative
solar-core radius. The blue curve shows the 5.49~MeV deuterium-capture
contribution; the red curve shows the
511~keV positron-annihilation contribution; and the black curve shows
their sum. Each contribution includes the primary monoenergetic line
and the broad lower-energy continuum populated by repeated Compton
scatterings.
}
\label{fig:Tot_Cascade}
\end{figure}
Furthermore, the pp reaction
\begin{equation}
  p + p \to d + e^{+} + \nu_{e}
  \label{eq:pp_reaction_intro}
\end{equation}
injects a positron that thermalizes and annihilates, yielding two quasi-monochromatic $511\;\mathrm{keV}$ photons that initiate their own cascade, extending the WISP spectrum further into the keV range.
These effects are illustrated in Fig.~\ref{fig:Tot_Cascade}, which shows how the primary 5.49 MeV and 511 keV injections are reprocessed into broad cumulative spectra extending to much lower energies.
\change{The relevance of solar $pp$ positrons as a source of exotic particles was
pointed out in Ref.~\cite{Gao:2025qkg}, which considered the direct annihilation
$e^+e^-\to\chi\bar\chi$ into millicharged particles. The mechanism discussed here
is different and complementary: we follow the \emph{real} 511~keV photons produced
in the dominant $\gamma\gamma$ channel and their subsequent reprocessing, which
applies to any WISP that mixes with the photon rather than only to states that can
be pair-produced from $e^+e^-$.} 
Since this reprocessed photon reservoir is independent of the specific WISP model, it provides a generic new source term for any photon-coupled WISP.
Both the Compton cascade and the 511 keV annihilation line have been overlooked in previous studies.\,\footnote{Although the relevance of pp positrons was noted in 
Ref.~\cite{OShea:2023gqn}, the conclusions drawn there were incorrect: those positrons thermalize before 
annihilating in the overwhelming majority of cases under solar-core 
conditions~\cite{ParticleDataGroup:2024cfk,Prantzos:2010wi}, so the 
annihilation produces a narrow line at $\simeq 511$~keV rather than 
the broad in-flight spectrum previously assumed.} 
As a result, the available photon source has been significantly underestimated.

The aim of this paper is to describe a general framework to quantify these new stellar WISP production channels. 
As a working example, we show specifically the case of the dark photon $A'$, the massive
gauge boson of a hidden $U(1)_D$ symmetry that couples to the visible
sector through kinetic mixing with the ordinary
photon~\cite{Holdom:1985ag,Fabbrichesi:2020wbt},
\begin{equation}
\mathcal{L} \supset -\frac{\chi}{2}\, F_{\mu\nu}F^{\prime \mu\nu}\,.
\label{eq:kinmix}
\end{equation}
The dark photon is among the best-studied WISPs, with a parameter space determined by only two quantities: the mass $m_{A'}$ and the kinetic-mixing parameter $\chi$.
This simplicity makes it an ideal benchmark and enables a largely model-independent assessment of the efficiency of the production mechanisms discussed above.
Furthermore, we restrict our analysis to the solar pp-chain, as the dominant source of high energy photons in the Sun. We provide more information about other solar channels in Appendix~\ref{app:solar_model}.

Our strategy, however, is more general and applies to any nuclear process—in the Sun or in other stars—that injects energetic photons or charged particles into the plasma, and to any WISP coupled to photons, including axions, dilatons, and chameleons, beyond the dark-photon case explicitly shown here. 
These energetic primaries can seed electromagnetic cascades in the plasma, thereby enhancing the production of weakly coupled light states.

\section{Compton cascade of non-thermal photons}
A non-thermal photon injected into the solar plasma at energy
$\omega_0 \gg T_\odot$ undergoes successive Compton scatterings off
ambient electrons, losing energy at each step until it is absorbed
into the thermal bath. In the presence of feebly interacting particles
that mix with the photon, each intermediate photon offers an
independent opportunity to convert into a WISP at its current energy,
generating a broad spectral continuum of conversion channels below
the injection line and enhancing the total WISP yield relative to
the primary-line estimate alone. \change{We describe this process by
first constructing the electromagnetic cascade and then weighting its
successive stages by the WISP conversion probability.}

\change{Compton scattering conserves photon number: it degrades the photon energy without removing the quantum, so that every intermediate photon remains available for conversion. This is what makes the cascade an efficient source. At high energies, only pair production could remove photons near the top of the cascade. However, this can be neglected due to the modest photon energy of interest here and the low atomic number of the solar plasma (dominated by hydrogen and helium). Quantitatively, the ratio of the pair-production rate to the Compton-scattering rate is $\Gamma_{\rm pair}/\Gamma_C \simeq 1.4\times 10^{-2}$ at the $5.49\;\MeV$ injection energy in the dominant production region, and vanishes below
the pair threshold. As their energy degrades, absorption through inverse bremsstrahlung becomes increasingly
important and ultimately terminates the cascade.}

\change{The short Compton mean free path in the solar core motivates a local
description: we neglect variations in the plasma properties along the
photon trajectory and evaluate the cascade at a fixed injection radius
$r$.} A Compton scattering off an electron at rest through an angle 
$\theta$ yields an outgoing photon of energy
\begin{equation}
\omega' = \frac{\omega}{1 + (\omega/m_e)(1-\cos\theta)}\,,
\label{eq:compton_kin}
\end{equation}
with $\omega_{\min}(\omega)\leq\omega'\leq\omega$, where 
$\omega_{\min}$ corresponds to back-scattering. 
The associated energy redistribution is encoded in the normalized Compton 
kernel~\cite{Younsi:2013gra}
\begin{equation}
K(\omega,\omega')
\equiv
\frac{1}{\sigma_{\rm KN}(\omega)}
\frac{\dd\sigma_{\rm KN}}{\dd\omega'}\,
\Theta(\omega' - \omega_{\min})\,\Theta(\omega - \omega')\,,
\label{eq:kernel}
\end{equation}
where $\dd\sigma_{\rm KN}/\dd\omega'$ and $\sigma_{\rm KN}(\omega)$ are
the Klein--Nishina differential and total cross section respectively.
By construction, $\int_{\omega_{\min}}^{\omega}\!\dd\omega'\;K(\omega,\omega') = 1$,
so that $K(\omega,\omega')$ gives the probability density for a photon
of energy $\omega$ to emerge with energy $\omega'$ after a single
scattering.

\change{Now, we have to take into account that at each stage, the photon either undergoes another Compton scattering or is absorbed through inverse bremsstrahlung. Since both rates are evaluated at the photon's current energy, the branching
between them is a local, energy-resolved statement: a photon of energy
$\omega$ at radius $r$ survives to scatter again with probability 
\begin{equation}
p(\omega,r)=\frac{\Gamma_C}{\Gamma_C+\Gamma_{\rm ff}}
=\frac{1}{1+\xi(\omega,r)}\,,
\qquad
\xi \equiv \frac{\Gamma_{\rm ff}}{\Gamma_C}\,,
\label{eq:survival}
\end{equation}
with $\Gamma_C$ and $\Gamma_{\rm ff}$ the Compton and inverse-bremsstrahlung
rates defined in Appendix~\ref{sec:cascade_duration}. Because
$\Gamma_{\rm ff}\propto\omega^{-3}$, this branching is strongly energy
dependent: absorption does not simply rescale each generation, it
\emph{reshapes} it, preferentially removing its low-energy tail. }

\change{To account for this effect, we define $f_n(\omega,r)$, with $n\geq 0$, as the energy distribution after $n$ Compton scatterings,
weighted by the probability of reaching that stage. The initial
distribution $f_0$ is normalized to unity; for a monoenergetic line,
$f_0(\omega,r)=\delta(\omega-\omega_0)$. Subsequent generations are defined as
\begin{equation}
f_{n+1}(\omega',r)=\int\!\dd\omega\;f_n(\omega)\,p(\omega,r)\,K(\omega,\omega')\,,
\label{eq:fn_iter}
\end{equation}
where the product $p(\omega,r)K(\omega,\omega')$ accounts
for both the probability of another scattering and its energy redistribution.
The cumulative
spectral weight is then
\begin{equation}
\mathcal{F}(\omega\change{,r})
\equiv \change{\sum_{n\ge0} f_n(\omega,r)}\,,
\label{eq:cumulative_F}
\end{equation}
which sums the contributions of the primary photon ($n=0$) and of all
subsequent cascade generations. The series converges for any $\xi>0$ and can
be resummed in closed form;
the
numerical implementation, together with the properties of $\mathcal{F}$, is
given in Appendix~\ref{sec:cascade_duration}.}

\change{We now include WISP production by assigning a conversion probability
$P(\omega_m,r_m)$ to each cascade stage, where $\omega_m$ and radius $r_m\simeq r$ are the photon energy and radial position at step $m$.\,\footnote{In general, $P$ may also
depend on other parameters, e.g. plasma conditions or specific
couplings related to the WISP considered.}
The details of the specific WISP model
enter \textit{only} through $P$. 
In the regime relevant here, $P\ll1$, so that different cascade steps add linearly up to corrections of order $P^2$.\,\footnote{For a fixed cascade history the exact probability is
$1-\prod_m[1-P(\omega_m,r_m)]$, which reduces to
$\sum_m P(\omega_m,r_m)$ when $P\ll1$.}
We stress that here $P$ is the probability per cascade step, so that each step of the cascade supplies exactly one conversion opportunity.
Retaining only the primary photon, as done in previous treatments of nuclear-line WISP production~\cite{Borexino:2012guz,Lucente:2022esm,DEramo:2023buu}, is therefore equivalent to truncating the cascade at the first interaction with the plasma.}

\change{The electromagnetic cascade and the conversion physics are thus
encoded in separate factors: $\mathcal{F}$ depends only on the
injection spectrum, the Compton kernel, and the local plasma
conditions, while all WISP-model dependence enters through $P$. The integral of $\mathcal{F}$ counts the conversion
opportunities available to a single injected photon. Averaged over the
solar source distribution we find
\begin{equation}
\int\!\dd\omega\,\mathcal{F}_{5.49}=31.3\,,
\qquad
\int\!\dd\omega\,\mathcal{F}_{511}=29.5\,,
\label{eq:F_integrals}
\end{equation}
i.e. about 90 conversion opportunities per $pp$ reaction once the three
injected photons are counted together (see below).}

With these definitions, the differential WISP emissivity sourced by
the cascade is
\begin{equation}
\frac{\dd\dot{n}_X^{\rm cas}}{\dd\omega}(\omega,r)
=
P(\omega,r)\,
\frac{\dd\dot n_\gamma^{\rm cas}}{\dd\omega}(\omega,r),
\label{eq:emissivity_compact}
\end{equation}
where
\begin{equation}
\frac{\dd\dot n_\gamma^{\rm cas}}{\dd\omega}
=
\dot{n}_\gamma^{\rm line}(r)\,
\mathcal{F}(\omega\change{,r})
\label{eq:line_injections}
\end{equation}
is the local differential photon source generated by the cascade. 
Here $\dot{n}_\gamma^{\rm line}(r)$ is the volumetric production rate of
photons in the nuclear line.
The $n=0$ term in $\mathcal{F}$ recovers the direct conversion of the
primary line photon, while the $n\geq1$ terms encode the cascade
enhancement. 
Since the $\gamma \to X$ conversion does not alter the energy of the
quantum, the functions $f_n$ describe equally the energy distributions
of the parent photons and of the resulting WISPs at each cascade stage.


The differential WISP flux at Earth follows by integrating the
emissivity over the solar volume,
\begin{equation}
\frac{\dd\Phi_X}{\dd\omega}
= \frac{1}{4\pi d_\odot^2}
\int_\odot\!\dd V\;
\frac{\dd\dot{n}_X^{\rm cas}}{\dd\omega}(\omega,r)\,,
\label{eq:flux_cascade}
\end{equation}
where $d_\odot \simeq 1\AU$ is the Earth--Sun distance.
This expression, together with the appropriate $P(\omega,r)$, is the
master formula for the cascade contribution to the solar WISP flux.
The factorized structure separates the cascade physics, encoded in
$\mathcal F(\omega\change{,r})$, from the particle physics, encoded in
$P(\omega,r)$. \change{Note that both factors depend on the radius, so
neither can be taken out of the volume integral; in practice
$\mathcal{F}$ is solved independently at each radius, which is legitimate
because the Compton mean free path in the solar core ($\sim0.04$--$0.3$~cm)
is extremely smaller than the scale over which the plasma varies.}

\section{Non-thermal sources in the solar pp chain}
%
%
We now specialize to the solar pp chain. In the notation of
Eq.~\eqref{eq:line_injections}, the relevant line rates
$\dot n_\gamma^{\rm line}(r)$ are those of the 5.49~MeV
deuterium-capture line and the 511~keV positron-annihilation line.
Since deuterons are injected by both the $pp$ and $pep$ reactions, the
local production rate of 5.49~MeV photons is
\begin{equation}
    \dot{n}_{5.49}(r)=\dot{n}_{pp}(r)+\dot{n}_{pep}(r)\,,
\end{equation}
where $\dot{n}_{pp}(r)$ and $\dot{n}_{pep}(r)$ denote the local
volumetric rates of the $pp$ and $pep$ reactions, respectively (see Appendix~\ref{app:solar_model} for details).
Similarly, positrons produced in the $pp$ reaction thermalize and
annihilate at rest, giving two photons per annihilation. Thus
\begin{equation}
    \dot n_{511}(r)\simeq 2\dot n_{pp}(r)\,,
\end{equation}
where the factor of 2 accounts for the two annihilation photons.


Thus, the total non-thermal WISP flux at Earth is given by the sum of the two contributions, and can be calculated starting from Eqs.~\eqref{eq:emissivity_compact} and~\eqref{eq:flux_cascade}, with 
\begin{align}
\frac{\dd\dot{n}_\gamma^{\rm cas}}{\dd\omega} =\;&
\mathcal{F}_{5.49}(\omega\change{,r})\,
\bigl(\dot n_{pp}(r)+\dot n_{pep}(r)\bigr)
\nonumber\\
&+2\,\mathcal{F}_{511}(\omega\change{,r})\,\dot n_{pp}(r)\,.
\label{eq:X_flux_total}
\end{align}
Here $\mathcal{F}_{5.49}$ and $\mathcal{F}_{511}$ are the cumulative spectral weights associated with the two injection energies. The first term in brackets describes the 5.49~MeV nuclear line and its cascade, sourced by deuteron production and therefore proportional to $\dot n_{pp}(r)+\dot n_{pep}(r)$. The second term describes the 511~keV positron-annihilation line and its cascade, sourced by the $pp$ reaction alone.\,\footnote{In the numerical implementation, the local rates are reconstructed from the normalized radial generation distributions provided by the solar model~\cite{Herrera:2023b23}; see Appendix~\ref{app:solar_model} for details.}
The local photon source spectrum in Eq.~\eqref{eq:X_flux_total},
evaluated at $r=0.005R_\odot$, is shown in Fig.~\ref{fig:Tot_Cascade}.

\section{Illustrative example: dark-photon production from the solar pp chain}
As a concrete example of application, we now consider the dark photon case.
The relevance of the pp-chain channel for dark photons was recently
emphasized by D'Eramo \emph{et al.}~\cite{DEramo:2023buu}, who pointed
out that the 5.49~MeV line can produce a sizeable flux of solar dark
photons through in-medium kinetic mixing and studied the associated
sensitivity of terrestrial experiments.
Their analysis, however, retained only the primary 5.49~MeV photon
injected by the nuclear reaction.
In the language introduced above, this corresponds to keeping only the
initial injection ($n=0$) and neglecting the subsequent electromagnetic
degradation of the photon spectrum.


\begin{figure}
    \centering
    \includegraphics[width=1\linewidth]{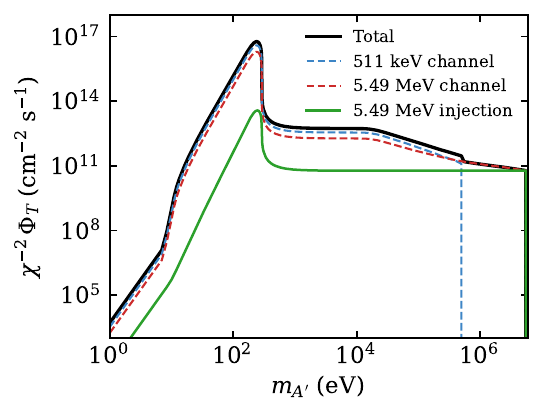}
    \caption{Total integrated transverse dark-photon number flux at Earth from pp-chain sources as a function of the mass (black), compared with the contribution of the 5.49~MeV primary line alone in the same realistic solar model (green).
    This last is the realistic-solar-model counterpart of the one-zone estimate of Ref.~\cite{DEramo:2023buu}. The contributions of the 5.49~MeV (dashed red) and 511~keV (dashed blue) channels, including their Compton cascades, are also shown.}    
    \label{fig:total_flux_DP}
\end{figure}

To apply our framework to dark photons, we must specify the photon--dark-photon
conversion probability $P(\omega,r)$. In a stellar plasma, the photon acquires
an in-medium dispersion relation through its self-energy, which modifies the
kinetic mixing with the dark photon~\cite{An:2013yfc,Redondo:2013lna,Hardy:2016kme}.
For the transverse polarization---which dominates the production at the
energies relevant here~\cite{DEramo:2023buu}---the conversion probability can be written
as~\cite{An:2013yfc,Redondo:2008aa,DEramo:2023buu}
\begin{equation}
P_T(\omega,r)
= \frac{\chi^2\,m_{A'}^4}
{\bigl(m_{A'}^2-\omega_{\rm pl}^2(r)\bigr)^2
+\bigl(\mathrm{Im}[\pi_T]\bigr)^2}\,,
\label{eq:P_T}
\end{equation}
where
$\omega_{\rm pl}(r)=\left(4\pi\alpha\,n_e(r)/m_e\right)^{1/2}$
is the local plasma frequency. 
Across the solar interior, it ranges from $\omega_{\rm pl}\sim 300~\mathrm{eV}$ in the core to negligible values near the surface.
The imaginary part of the transverse
photon self-energy is \change{set by the total damping rate of the photon
field}~\cite{An:2013yfc,Redondo:2008aa,DEramo:2023buu},
\begin{equation}
\mathrm{Im}[\pi_T]
= -\,\omega\,\Gamma_{\rm tot}(\omega,r)\,,
\qquad
\Gamma_{\rm tot}=\Gamma_C+\Gamma_{\rm ff}\,,
\label{eq:Im_pi}
\end{equation}
\change{where $\Gamma_C=n_e\,\sigma_{\rm KN}(\omega)\bigl(1-e^{-\omega/T}\bigr)$, with $\sigma_{\rm KN}$ the full Klein--Nishina cross section,
is the Compton damping rate and $\Gamma_{\rm ff}$ the inverse-bremsstrahlung absorption rate. 
Compton scattering dominates at the injection energies, while free-free absorption takes over below $\simeq9$~keV.}
For the photon energies of interest, $\omega\gg T$, the Bose factor
$\bigl(1-e^{-\omega/T}\bigr)$ is effectively unity.  

Equation~\eqref{eq:P_T} reduces to the screened result 
$P_T \simeq \chi^2 m_{A'}^4/\omega_{\rm pl}^4$ for 
$m_{A'}\ll\omega_{\rm pl}$, and to the vacuum value 
$P_T\simeq\chi^2$ for $m_{A'}\gg\omega_{\rm pl}$. 
The phenomenologically richest case is resonance, $m_{A'}^2 = \omega_{\rm pl}^2(r_{\rm res})$, where the real 
part of the denominator vanishes and conversion occurs in a narrow radial shell whose position depends on $m_{A'}$ through the profile of $\omega_{\rm pl}(r)$. In this regime, $P_T$ reduces to
$P_T^{\rm res} = \chi^2 m_{A'}^4/(\mathrm{Im}[\pi_T])^2$\change{. The quantity
that enters the flux, however, is the integral of $P_T$ across the shell,
whose width is itself controlled by $\mathrm{Im}[\pi_T]$. The integrand is a
Lorentzian in $\omega_{\rm pl}^2$ and the integral can be done analytically
(see Appendix~\ref{sec:resonance}),
\begin{equation}
\int\!\dd r\,P_T\Big|_{\rm res}
=\frac{\pi\,\chi^2 m_{A'}^4}
{\bigl|\mathrm{Im}[\pi_T]\bigr|\;
\bigl|\dd\omega_{\rm pl}^2/\dd r\bigr|}\Bigg|_{r_{\rm res}}
\;\propto\;\frac{1}{\omega\,\Gamma_{\rm tot}(\omega)}\,,
\label{eq:res_integral}
\end{equation}
so that one power of the damping rate is recovered by the radial integration.}

With the formalism developed here, it is straightforward to extend the
previous result~\cite{DEramo:2023buu} beyond the primary 5.49~MeV line:
one inserts the dark-photon conversion probability $P_T(\omega,r)$ into
the general flux formula, Eq.~\eqref{eq:X_flux_total}, thereby including
both non-thermal injection channels and their Compton cascades. The
result is shown in Fig.~\ref{fig:total_flux_DP}.

Away from resonance, $P_T$ is approximately independent of $\omega$.
The cascade then enters mainly as a multiplicative counting factor: the
integrated flux is enhanced by \change{a factor $\simeq90$, the number of}
conversion opportunities
contained in the cumulative cascade weights---one cascade seeded by the
5.49~MeV line and two by the 511~keV annihilation photons---relative to
the single primary 5.49~MeV opportunity retained in
Ref.~\cite{DEramo:2023buu}.
In the resonant region, \change{$m_{A'}\simeq 100$--$290$~eV}, the enhancement is
much larger, \change{reaching about three orders of magnitude}. The reason is that
\change{$\int\dd r\,P_T^{\rm res}\propto1/[\omega\Gamma_{\rm tot}(\omega)]$}, so lower-energy photons convert with
parametrically larger probability. The Compton cascade transfers
spectral weight from the original MeV- and 511~keV-scale injections to
much lower energies, where the resonant conversion probability is
larger, amplifying the flux well beyond the constant-$P_T$ counting.
\change{The lower end of the quoted resonant range is set by the requirement
that the resonant shell still overlaps the region where the nuclear
injections occur: since $\omega_{\rm pl}$ falls monotonically outwards, at
$m_{A'}=100$~eV the shell already sits at $r\simeq0.29\,R_\odot$, where the
$pp$ emissivity has dropped to $2\%$ of its peak and $99.5\%$ of the
injections have taken place further in. The upper end is
$\omega_{\rm pl}(0)=289$~eV, above which no resonance exists anywhere in the
Sun.}

\change{Finally, we ask where in the spectrum this non-thermal flux actually
dominates. The relevant comparison is not with the primary-line estimate but
with the thermal solar plasma, and it is independent of the WISP model: the
conversion probability multiplies both reservoirs and cancels from the ratio,
so what matters is the ratio of the two photon source spectra,
$\dd\dot n_\gamma^{\rm cas}/\dd\omega$ from Eq.~\eqref{eq:X_flux_total} and
$\Gamma_{\rm tot}\,\dd n_\gamma^{\rm BB}/\dd\omega$ for the bath, with
$\dd n_\gamma^{\rm BB}/\dd\omega=\omega^2/[\pi^2(e^{\omega/T}-1)]$. Integrating
both over the solar volume, we find that the non-thermal cascade overtakes the
thermal reservoir at
\begin{equation}
\omega_\times \simeq 89\;\keV \,,
\label{eq:crossover}
\end{equation}
and, because the thermal spectrum is Boltzmann suppressed, the margin grows to more
than thirty orders of magnitude within one decade in energy. Above
$\sim100\;\keV$, therefore, essentially every solar dark photon originates from
the non-thermal channels described here.}

\begin{figure}[t]
    \centering
    \includegraphics[width=1\linewidth]{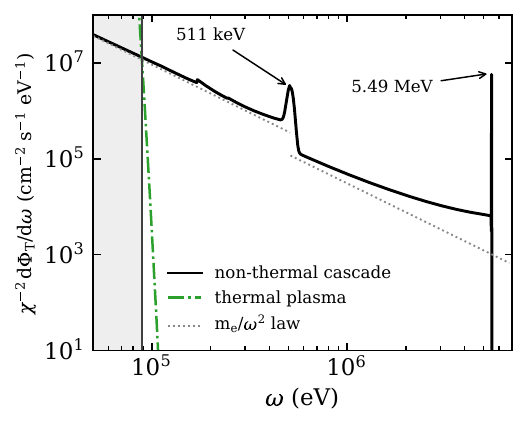}
    \caption{\change{Differential solar dark-photon flux at Earth in the
    non-thermal window, for $m_{A'}$ large enough that $P_T\simeq\chi^2$
    (black), compared with the flux from the thermal plasma (green
    dash-dotted). The vertical line marks the crossover of
    Eq.~\eqref{eq:crossover}, $\omega_\times\simeq89$~keV; the thermal
    reservoir dominates in the shaded region and is exponentially
    suppressed above it. The dotted line is the universal cascade law
    $\mathcal{F}\simeq m_e/\omega^2$, normalized to
    $3\Phi_{pp}$ below the annihilation line and to
    $\Phi_{pp}+\Phi_{pep}$ above it.
    The asymptotic scaling fails near the 5.49 MeV line, where only the first few scatterings contribute.
    Both lines are shown with their thermal Doppler widths.}}
    \label{fig:window}
\end{figure}

\change{Figure~\ref{fig:window} shows the resulting flux in that window. Three
features characterize it. First, a continuum approaching the universal cascade law $\mathcal{F}\simeq m_e/\omega^2$ derived in
Appendix~\ref{sec:cascade_duration}, which is independent of the injection
energy and of the WISP species. Second, a step of a factor of three across
511~keV: below the annihilation line all three photons injected per $pp$
reaction contribute to the continuum, whereas above it only the 5.49~MeV
cascade survives. Third, the two lines themselves, which are not delta
functions but are thermally broadened by the motion of the emitting system,
$\sigma_E/E=\sqrt{T/M}$ with $M$ its total mass; this gives
$\sigma_E\simeq16.5\;\keV$ for the 511~keV line, where the emitter is the
annihilating pair ($M=2m_e$), against $\simeq4\;\keV$ for the 5.49~MeV line,
where it is the much heavier $p+d$ system.}

\change{Counting only this window, the non-thermal reservoir provides 19.7
conversion opportunities per $pp$ reaction above $100\;\keV$, against the single
opportunity retained in Ref.~\cite{DEramo:2023buu}. 
The statement is sharper still below the
injection line: in the decade $0.1$--$1\;\MeV$ a treatment retaining only the
primary photons predicts \emph{no} flux whatsoever, since the only photon it
contains sits at $5.49\;\MeV$, whereas the reprocessed reservoir supplies $17.7$
conversion opportunities per $pp$ reaction, carried by the $511\;\keV$ line and
the continuum around it. In that decade the reprocessed photons are not just an
enhancement of a known signal but rather the entire signal. However, for an experiment sensitive to the $2$--$6\,\MeV$ band, such as the underground liquid-scintillator detector JUNO~\cite{JUNO:2015zny,JUNO:2020hqc}, the reservoir provides $1.6$ conversion opportunities per $pp$ reaction. Therefore, the corresponding improvement in the projected limits of Refs.~\cite{Lucente:2022esm,DEramo:2023buu} is negligible.}

\section{Discussion and conclusions}
We have presented a general framework for computing stellar WISP
production from non-thermal nuclear injections. Two ingredients,
both incorporated here for the first time, drive the result: the
full Compton cascade of the primary photons, which generates a broad
spectral continuum of conversion opportunities below each injection
line, and the 511\,keV positron-annihilation line, which we identify
as \change{a} previously overlooked source of non-thermal
WISP production and which seeds its own cascade extending the
spectrum further into the keV range.
The formalism is built around the cumulative spectral weight
$\mathcal{F}(\omega\change{,r})$, which encodes the cascade physics independently
of the WISP model, and the conversion probability $P(\omega,r)$,
which encodes the particle physics. This factorized structure makes
the framework directly applicable to any WISP that mixes with the
photon---including axion-like particles, dark photons, dilatons, and
chameleons---and to any stellar environment hosting nuclear reactions
that inject energetic photons or positrons.

As a concrete application, we have worked out the dark photon flux
from the solar pp chain using a realistic solar model. The combined
inclusion of the Compton shoulder and the 511\,keV line increases
the total integrated dark photon flux by \change{about two orders of magnitude
off resonance and by three orders of magnitude in the resonant band,}
relative to previous estimates based on the primary lines alone.
\change{More importantly, it changes the shape of the predicted spectrum in a
way that is directly relevant to experiment. Above $\simeq89$~keV the thermal
solar plasma is exponentially dead, so the reprocessed photons are the only
source of solar dark photons; that window carries an $\omega^{-2}$ continuum,
a factor-of-three step at the annihilation line, and two lines of calculable thermal width. 
Since the cumulative weight
$\mathcal{F}$ is a purely electromagnetic quantity, the same window and the
same spectral shape apply to any photon-coupled WISP.}

The non-thermal fluxes computed here can, in principle, be searched
by a variety of experiments, for example in underground detectors
sensitive to WISP absorption~\cite{XENON:2024wpa,Yuan:2025twx} or in
helioscopes such as the forthcoming IAXO~\cite{IAXO:2019mpb,IAXO:2020wwp,IAXO:2024wss}.
The actual experimental sensitivity will depend on the specific WISP
species, the energy range accessible to each experiment, and the
details of the detector response. A thorough assessment of the
discovery potential would therefore require a dedicated analysis
tailored to specific WISPs and experimental designs, which lies
beyond the scope of this paper.

\acknowledgements

\textbf{\textit{Acknowledgments.}}---This article is based upon work from COST Action COSMIC WISPers (CA21106).\\
M.~Giannotti acknowledges support from the Spanish Agencia Estatal de Investigación through grant PID2019-108122GB-C31, funded by MCIN/AEI/10.13039/501100011033, and from the European Union NextGenerationEU/PRTR initiative (Planes Complementarios, Programa de Astrofísica y Física de Altas Energías). M.~Giannotti also acknowledges support from grant PGC2022-126078NB-C21, ``Aún más allá de los modelos estándar,'' funded by MCIN/AEI/10.13039/501100011033 and by the European Regional Development Fund (ERDF), ``A way of making Europe''; from the European Union's Horizon 2020 research and innovation programme under European Research Council (ERC) grant agreement No.\ 788781 (IAXO+); and from project CNS2025-165965, ``Señales de axiones en el rango MeV desde fuentes astrofísicas hasta detectores actuales y futuros,'' funded by MICIU/AEI/10.13039/501100011033. 
G.L. acknowledges support from the U.S. Department of Energy under contract number DE-AC02-76SF00515. V.F. acknowledges support from the Universidad de Zaragoza under the predoctoral contract call PI-PRD/2024-001. 

\appendix

In these appendices we provide additional details of the solar model used in our numerical analysis, as well as technical aspects underlying the results presented in the main text.

\section{The solar model}
\label{app:solar_model}

In this work we have used the AAG21/SF3 Standard Solar Model from the Serenelli data release~\cite{Herrera:2023b23}. 
The package provides both the solar structure profiles and the neutrino-source information
needed for our calculation. 
In particular, the \texttt{struct+nu\_}
files contain radial profiles of the solar structure variables together
with the normalized neutrino-generation distributions for the sources
$pp$, $pep$, $hep$, ${}^{7}\mathrm{Be}$, ${}^{8}\mathrm{B}$,
${}^{13}\mathrm{N}$, ${}^{15}\mathrm{O}$, and ${}^{17}\mathrm{F}$,
while the \texttt{fluxes\_} files provide the corresponding neutrino
fluxes~\cite{Herrera:2023b23}.

\begin{figure*}[t!]
    \centering
    \includegraphics[width=1\linewidth]{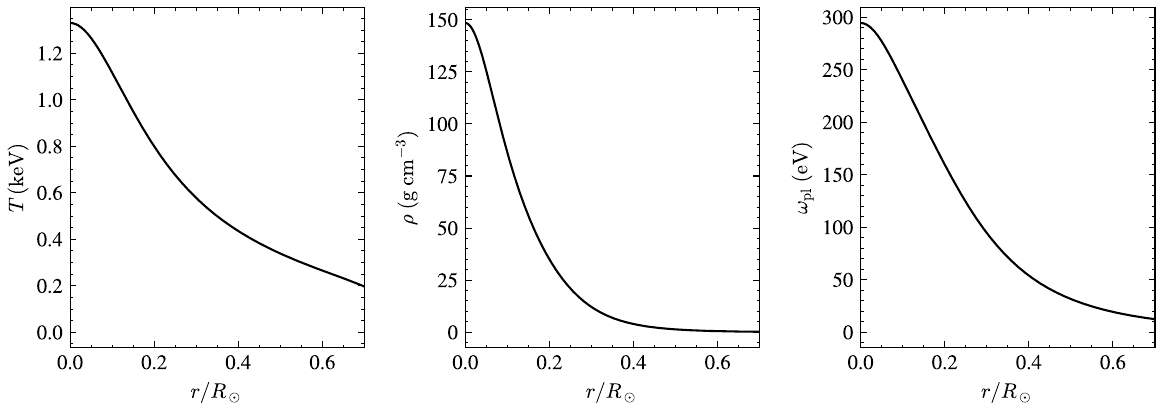}
    \caption{Profile of temperature, density and plasma frequency of the solar model~\cite{Herrera:2023b23} adopted in our work.}
    \label{fig:Solar_profiles}
\end{figure*}

Figure~\ref{fig:Solar_profiles} shows the radial profiles of temperature $T$ (left panel), density $\rho$ (middle panel) and plasma frequency $\omega_{\rm pl}$ (right panel) adopted in our work. These quantities are peaked at the center of the Sun and then monotonically decrease with the radius $r$.

We employ the \texttt{fluxes\_SF3\_AAG21.dat} file to compute the local volumetric production rates $\dot n_i(r)$ used in the main text. First, we infer the total solar production rate for each source $i$ from the tabulated neutrino flux at Earth ($\Phi_i$) via
\begin{equation}
\dot N_i^{\rm tot}=4\pi d_\odot^2\,\Phi_i\,,
\label{eq:Ntot_from_flux}
\end{equation}
where $d_\odot=1\,\mathrm{AU}$.\,\footnote{Although the README file associated with the model describes the \texttt{fluxes\_} files as 
``number per second'', the numerical values correspond to the standard
solar neutrino fluxes at Earth, so the conversion in
Eq.~\eqref{eq:Ntot_from_flux} is required.}.
To connect $\dot N_i^{\rm tot}$ to $\dot n_i$, it is convenient to introduce the dimensionless
radius
\begin{equation}
R \equiv \frac{r}{R_\odot}\,,
\end{equation}
where $R_\odot=6.96\times 10^{10}$~cm is the solar radius.
The solar model~\cite{Herrera:2023b23} tabulates, for each source $i$, a normalized radial
generation profile in the corresponding \texttt{nu\_}$i$ column. In what
follows, we denote this profile by $\nu_i(R)$, so that
\begin{equation}
\int_0^1 \nu_i(R)\,\dd R = 1\,.
\label{eq:nu_norm}
\end{equation}
Thus, $\nu_i(R)$ is the shell-weighted radial distribution associated with source $i$, implying that the production rate in the shell
$(R,R+\dd R)$ is
\begin{equation}
\dd \dot N_i
=
\dot N_i^{\rm tot}\,\nu_i(R)\,\dd R\,,
\label{eq:dNi_shell}
\end{equation}
or equivalently
\begin{equation}
\frac{\dd \dot N_i}{\dd r}
=
\frac{\dot N_i^{\rm tot}}{R_\odot}\,\nu_i(R)\,.
\label{eq:dNidr_shell}
\end{equation}
Therefore, the corresponding local volumetric production rate is
\begin{equation}
\dot n_i(r)
=
\frac{1}{4\pi r^2}\frac{\dd \dot N_i}{\dd r}
=
\frac{\dot N_i^{\rm tot}}{4\pi r^2 R_\odot}\,\nu_i(R)\,.
\label{eq:ni_local}
\end{equation}

Using these ingredients, we could reconstruct the non-thermal sources relevant for the
present work as follows:
\begin{itemize}
    \item The injected photons at 5.49~MeV originate from
\begin{equation}
p+d\to{}^{3}\mathrm{He}+\gamma\,.
\end{equation}
Since deuterons are injected by both the $pp$ and $pep$ reactions, the
corresponding shell source is
\begin{equation}
\dd \dot N_{5.49}
=
\Big[
\dot N_{pp}^{\rm tot}\,\nu_{pp}(R)
+
\dot N_{pep}^{\rm tot}\,\nu_{pep}(R)
\Big]\dd R\,.
\label{eq:dN549_shell}
\end{equation}
Equivalently,
\begin{equation}
\dot n_{5.49}(r)
=
\dot n_{pp}(r)+\dot n_{pep}(r)\,.
\label{eq:n549_local}
\end{equation}
Thus, the integrated 5.49~MeV photon source, which is not tabulated directly in the solar
model but inferred from the neutrino-source bookkeeping, is $\dot N_{5.49}=1.691\times 10^{38}$\,s.

\item The dominant positron source relevant for the present calculation is the
$pp$ reaction,
\begin{equation}
p+p\to d+e^+ + \nu_e\,,
\end{equation}
which injects one positron per reaction. Therefore,
\begin{equation}
\dd \dot N_{e^+}
=
\dot N_{pp}^{\rm tot}\,\nu_{pp}(R)\,\dd R\,,
\label{eq:dNep_shell}
\end{equation}
or equivalently
\begin{equation}
\dot n_{e^+}(r)\simeq \dot n_{pp}(r)\,.
\label{eq:nep_local}
\end{equation}
More generally, the integrated positron production $\dot N_{e^+}=1.698\times 10^{38}$\,s$^{-1}$ receives direct
contributions from all $\beta^+$-producing channels tabulated by the
solar model, namely $pp$, $hep$, ${}^{8}\mathrm{B}$,
${}^{13}\mathrm{N}$, ${}^{15}\mathrm{O}$, and ${}^{17}\mathrm{F}$. However, the
sum of the subdominant channels changes the \(pp\) result by less than
\(1\%\).

As discussed in the following section, to a very good approximation all $pp$ positrons thermalize
before annihilation. 
Their annihilation therefore produces a narrow 511~keV line, with two photons per positron. 
The corresponding shell source is
\begin{equation}
\dd \dot N_{511}
=
2\,\dot N_{pp}^{\rm tot}\,\nu_{pp}(R)\,\dd R\,,
\label{eq:dN511_shell}
\end{equation}
or equivalently
\begin{equation}
\dot n_{511}(r)\simeq 2\,\dot n_{pp}(r)\,.
\label{eq:n511_local}
\end{equation}
The factor of 2 appears because the spectral function
$\mathcal{F}_{511}(\omega,r)$ is normalized per injected photon.
\end{itemize}

We note that the AAG21/SF3 solar model~\cite{Herrera:2023b23} also provides data for additional monochromatic $\gamma$-ray injections from other reactions in the pp chain and the CNO cycle. In this work we consider only the dominant deuterium-capture line $p+d\to{}^{3}\mathrm{He}+\gamma$ at $5.49$ MeV and neglect all other nuclear lines. The largest omitted contribution is
${}^{3}\mathrm{He}+{}^{4}\mathrm{He}\to{}^{7}\mathrm{Be}+\gamma$, which produces a $1.59$ MeV photon with total rate $\dot N_{1.59}=1.271\times10^{37}\,{\rm s^{-1}}$, about an order of magnitude below $\dot N_{5.49}$. All remaining lines are even smaller and are likewise neglected. 

\begin{figure}
    \centering
    \includegraphics[width=0.75\linewidth]{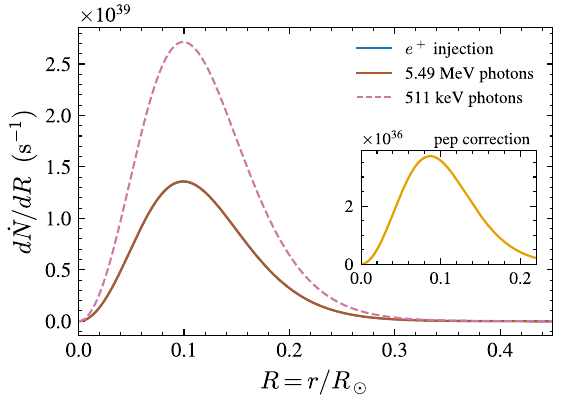}
    \caption{Radial shell injection rates relevant for the non-thermal cascade: 5.49 MeV photons from deuterium capture (orange), positrons from pp reactions (blue), and the corresponding 511 keV annihilation photons (dashed purple). The 5.49 MeV source almost overlaps the positron profile (hence, it is not visible in the plot) because it is dominated by pp reactions; the inset shows the smaller pep contribution separately. }
    \label{fig:Shell_injectionrates}
\end{figure}

Figure~\ref{fig:Shell_injectionrates} shows the radial shell injection rates that seed the non-thermal cascade.
In the main panel, the 5.49~MeV curve sits on top of the positron curve and hides it from view, since both are dominated by the $pp$ reaction; the small pep contribution to deuteron production---the only difference between them---is shown separately in the inset.
The 511~keV annihilation-photon source follows the same radial profile as the positron source, with twice the normalization, since each thermalized positron annihilates into two photons (in-flight annihilation is negligible, as shown in the section below).
Both non-thermal photon sources---the 5.49 MeV line and the 511 keV line---are thus produced in the same central region of the Sun, peaking around $r\simeq 0.1\,R_\odot$, \change{the radius we use as a reference when quoting local quantities in
Appendix~\ref{sec:cascade_duration}}.

\section{Thermalization of positrons in the Sun}
\label{sec:annihilation}
The pp chain produces a positron in the reaction $p + p \rightarrow d + e^+ + \nu_e$, which has a $Q$-value $Q=2 m_p-m_d-m_e \simeq$ 420 keV . Because the colliding protons are essentially at rest at solar-core temperatures and the deuteron carries negligible recoil kinetic energy, this $Q$-value is shared between the positron and the neutrino, $K_{e^{+}}+E_\nu \simeq Q$, with $K_{e^{+}}$ distributed continuously up to the endpoint $K_{e^{+}}^{\max } \simeq Q$. In what follows we take this endpoint as a conservative initial energy for the positron: it maximizes the slowingdown path and thus the in-flight annihilation probability, providing an upper bound on the fraction of pp positrons that annihilate before thermalizing.

As they travel, positrons will interact via the electromagnetic force with the much slower solar plasma components, losing energy in the interaction. The rate of energy losses depends on the initial energy of the positron, $K_{e^+}$, and the densities of the target particles. 

In the energy regime of interest, this occurs mainly via Coulomb scattering. Bremsstrahlung only dominates at much higher velocities and ionization is way subdominant. For highly energetic positrons, the target electrons can be considered at rest and the energy loss rate depends essentially on the number density of electrons as \cite{Prantzos:2010wi}
\begin{widetext}
\begin{equation}\label{eq: COU loss}
    \left(\frac{\text{d}E}{\text{d}t}\right)_C = - \frac{7.7\times10^{-9}}{v}\left(\frac{n_e}{\text{cm}^{-3}}\right)\Big[\ln\gamma -\ln\left(\frac{n_e}{\text{cm}^{-3}}\right)+73.6\Big] \frac{\eV}{\text{s}}.
\end{equation}
\end{widetext}

In the expression, $v$ is the positron's speed and $\gamma$ its Lorentz factor. As positrons slow down, a significant fraction of them may annihilate in flight, reducing the number of WISPs produced in the $511\;\keV$ line. The probability that a positron with initial kinetic energy $K_0$ anihilates before reaching energy $K$ is given by
\begin{widetext}
\begin{equation} \label{eq: P in fligt}
    P_{\text{in-flight}}(K_0,K) = 1 - \exp\left(-n_e\int_K^{K_0}\frac{v(E')\sigma_{\text{ann}}(E')\; \text{d}E'}{|\frac{\text{d}E'}{dt}|}\right),
\end{equation}
\end{widetext}
where $\sigma_{\text{ann}}$ is the Heitler's annihilation cross-section \cite{GEANT4:2002zbu,geant4_prm_annihilation},
\begin{widetext}
\begin{equation}
    \sigma_{\text{ann}} = \frac{\pi r_e^2}{\gamma+1}\Bigg[\frac{\gamma^2+4\gamma+1}{\gamma^2-1}\ln\left(\gamma + \sqrt{\gamma^2-1}\right)-\frac{\gamma+3}{\sqrt{\gamma^2-1}}\Bigg]\,,
\end{equation}
\end{widetext}
being $r_e=2.8\times 10^{-13}\,{\rm cm}$ the classical electron radius.
For a positron produced with kinetic energy $K_{e^{+}}^{\max } \simeq Q$, the probability that it survives and thermalizes to $K^{\rm th}\sim 3\,\keV$\,\footnote{$K^{\rm th}$ denotes the thermalization energy scale in the Sun, of the order of the solar-core thermal energy. Its precise value does not affect our conclusions.} is given by $P_{\text{surv}} = 1-P_{\text{in-flight}}(420\,{\rm keV}, 3\,{\rm keV})$. We find that $P_{\text{surv}} \sim 0.97$ for the entire radius of the Sun. For any positron with a lower initial energy, the probability of surviving increases up to $\sim 0.99$. We can therefore assume that the number of positrons annihilating before thermalizing is negligible and thus all positrons produced in the pp chain thermalize and annihilate at thermal velocities, at a rate $\dot{n}_\text{ann} \simeq \dot{n}_{pp}$.

\section{\change{Absorption during the Compton cascade}}
\label{sec:cascade_duration}

As discussed in the main text, the non-thermal photons injected by nuclear reactions undergo repeated Compton scattering off solar plasma electrons, leading to progressive energy degradation and providing additional opportunities for conversion into WISPs.
This is the Compton cascade.

\change{The cascade does not continue indefinitely. At each step the photon
either scatters again, at a rate $\Gamma_C$, or is truly absorbed, at a rate
$\Gamma_{\rm ff}$; Eq.~\eqref{eq:survival} of the main text is the
branching ratio between the two. In the fully ionized solar core the dominant
absorption channel is inverse bremsstrahlung, whose rate is~\cite{rybicki_lightman_1979}}  
\begin{widetext}
\begin{equation}
\frac{\Gamma_{\mathrm{ff}}(\omega, r)}{\mathrm{eV}}=4.80 \times 10^{-42}\left(\frac{T}{\mathrm{eV}}\right)^{-1 / 2}\left(\frac{n_e}{\mathrm{~cm}^{-3}}\right) \sum_i Z_i^2\left(\frac{n_i}{\mathrm{~cm}^{-3}}\right)\left(\frac{\omega}{\mathrm{eV}}\right)^{-3}\left(1-e^{-\omega / T}\right) .
\end{equation}
\end{widetext}

\begin{figure*}[t]
    \centering
    \includegraphics[width=1\linewidth]{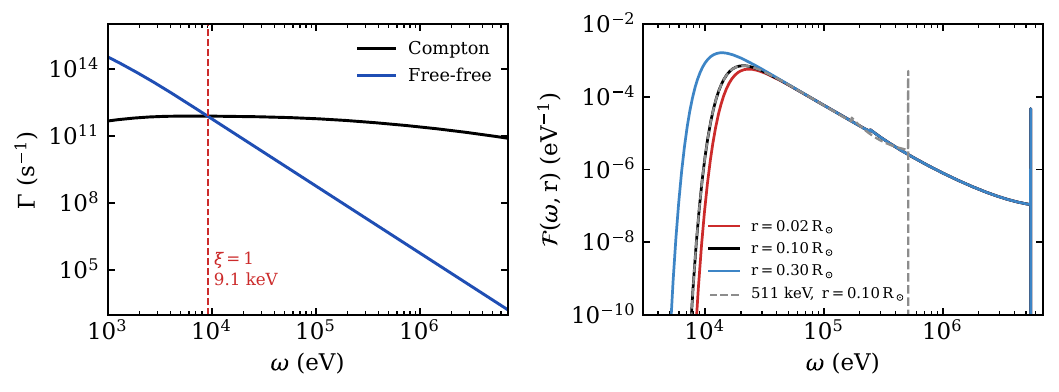}
    \caption{\change{\textit{Left:} Compton (black) and inverse-bremsstrahlung
    (blue) rates at $r=0.1\,R_\odot$ as a function of photon energy; the dashed
    line marks the energy at which $\xi=\Gamma_{\rm ff}/\Gamma_C$ crosses unity.
    \textit{Right:} the cumulative spectral weight
    $\mathcal{F}(\omega,r)$ of Eq.~\eqref{eq:resolvent} for the 5.49~MeV
    injection at three radii, and for the 511~keV injection at
    $r=0.1\,R_\odot$ (grey dashed). Above $\simeq30\;\keV$ the four curves lie
    on top of one another, illustrating that $\mathcal{F}$ is there
    independent both of the radius and of the injection energy; the radial
    dependence appears only in the low-energy tail, where absorption acts.}}
    \label{fig:cascade_absorption}
\end{figure*}

Here $n_i$ is the number density of nuclei with atomic number $Z_i$
(expressed in $\mathrm{cm}^{-3}$), $\omega$ is the photon energy in eV, $T$ is the temperature in $\text{eV}$, and the sum runs over all ionic species in the
solar plasma. \change{The Compton rate is
$\Gamma_C=n_e\,\sigma_{\rm KN}(\omega)\bigl(1-e^{-\omega/T}\bigr)$, evaluated
with the full Klein--Nishina cross section, so that the same $\Gamma_C$ enters
the branching ratio $\xi$ and the damping rate $\Gamma_{\rm tot}$ of
Eq.~\eqref{eq:Im_pi}. The two rates are shown in the left panel of
Fig.~\ref{fig:cascade_absorption}. Their ratio falls as $\omega^{-3}$ at high
energy and crosses unity at $\omega\simeq9.1\;\keV$, i.e.\ $8.1\,T$, at the
reference radius $r=0.1\,R_\odot$; the crossing moves to $10.1\;\keV$ at
$0.05\,R_\odot$ and to $7.1\;\keV$ at $0.2\,R_\odot$, so the cascade survives
to lower energies as the plasma thins outwards.}  

\change{\textit{Numerical implementation.}---We discretize the Compton kernel
of Eq.~\eqref{eq:kernel} on a logarithmic energy grid,
$M_{ij}=\int_{{\rm bin}\,j}\dd\omega'\,K(\omega_i,\omega')$, so that one cascade
generation is one matrix multiplication and photon number is conserved exactly,
$\sum_j M_{ij}=1$. The iteration of Eq.~\eqref{eq:fn_iter} then reads
$f_{n+1}=f_n\,D\,M$ with $D={\rm diag}\bigl[1/(1+\xi(\omega_i,r))\bigr]$, and the
cumulative weight of Eq.~\eqref{eq:cumulative_F} is
\begin{equation}
\mathcal{F}=f_0\,\bigl(\mathbf{1}-D\,M\bigr)^{-1}\,.
\label{eq:resolvent}
\end{equation}
The resummation is well defined because $D M$ has spectral radius below unity,
whereas $M$ by itself has spectral radius exactly one. 
In other words, the absorption renders the cascade finite.}

\change{We use 600 bins per decade between $300\;\eV$ and $7\;\MeV$; doubling the
resolution changes the number of conversion opportunities above $100\;\keV$ by
four parts in $10^{5}$. Equation~\eqref{eq:resolvent} is solved independently at
54 radii between $0.002$ and $0.85\,R_\odot$ and carried inside the volume
integral of Eq.~\eqref{eq:flux_cascade}.}

\change{\textit{Properties of the cumulative weight.}---Averaging the Compton
relation in Eq.~\eqref{eq:compton_kin} over one scattering gives
$\langle1/\omega\rangle_n-\langle1/\omega\rangle_{n-1}=\langle1-\cos\theta\rangle/m_e$,
which tends to $1/m_e$ once $\omega\ll m_e$.
Generation $n$ therefore sits at
$\omega_n\simeq m_e/n$, and the cumulative weight follows the universal law
\begin{equation}
\mathcal{F}(\omega,r)\;\simeq\;\frac{m_e}{\omega^{2}}\,,
\qquad
\omega_{\rm min}\lesssim\omega\ll\omega_0\,,
\label{eq:universal_F}
\end{equation}
independent of the injection energy, of the radius, and of the WISP species.
We find that Eq.~\eqref{eq:universal_F} reproduces the numerical behavior 
to within $10$--$20\%$ 
across the continuum; the law fails only within a
factor of a few of each injection line, where the shape is still set by the
first one or two scatterings. This is the reason the framework is genuinely
WISP-independent: $\mathcal{F}$ is fixed by Compton kinematics alone, and the
entire particle-physics content sits in $P(\omega,r)$.}

\change{Integrating Eq.~\eqref{eq:resolvent}, at $r=0.1\,R_\odot$ we obtain
$\int\dd\omega\,\mathcal{F}=30.2$ and $28.4$ for the 5.49~MeV and 511~keV
injections, and $31.3$ and $29.5$ once averaged over the solar source
distribution, which are the values quoted in Eq.~\eqref{eq:F_integrals}.}

\section{\change{Radial integration through the resonance}}
\label{sec:resonance}

\change{The volume integral of Eq.~\eqref{eq:flux_cascade} contains, for every
energy bin, the Lorentzian factor
$\bigl[(m_{A'}^2-\omega_{\rm pl}^2(r))^2+\mathrm{Im}[\pi_T]^2\bigr]^{-1}$. Near a
resonance this is a very narrow feature: its half width in radius is
\begin{equation}
\frac{\Delta r_{\rm res}}{R_\odot}
=\frac{\bigl|\mathrm{Im}[\pi_T]\bigr|/\omega_{\rm pl}^2}
{\bigl|\dd\ln\omega_{\rm pl}^2/\dd x\bigr|}\,,
\qquad x\equiv r/R_\odot\,,
\end{equation}
which at $r_{\rm res}=0.1\,R_\odot$ amounts to $0.6$ and $0.03$ cells of the
$5\times10^{-4}\,R_\odot$ grid on which the solar model is tabulated, for
$\omega=511\;\keV$ and $10\;\keV$ respectively. A quadrature performed directly
on that grid would therefore sample the resonance at essentially arbitrary
points, and we avoid it by integrating the Lorentzian in closed form.}

\change{Substituting $u=\omega_{\rm pl}^2(r)$, which is monotonic, and
$s=u-m_{A'}^2$, and taking on each panel $[x_k,x_{k+1}]$ the Jacobian
$\dd x/\dd u$ and the width $a\equiv\mathrm{Im}[\pi_T]$ as constant and the
source $S$ as linear in $u$, every remaining integral is elementary,
\begin{equation}
\int_{x_k}^{x_{k+1}}\!\!\dd x\,\frac{S}{s^2+a^2}
=\frac{\Delta x}{\Delta u}
\left[\bar S\,P_0+\frac{\Delta S}{\Delta u}\bigl(P_1-\bar s\,P_0\bigr)\right],
\end{equation}
with
$P_0=a^{-1}[\arctan(s_1/a)-\arctan(s_0/a)]$ and
$P_1=\tfrac12\ln[(s_1^2+a^2)/(s_0^2+a^2)]$. 
The scheme is uniformly valid: when the pole lies inside
a panel it reproduces the narrow-resonance result of
Eq.~\eqref{eq:res_integral}, with $S$ evaluated at the pole rather than at the
panel midpoint, and far from resonance it reduces to the ordinary midpoint rule.
The panel grid therefore has to resolve only the smooth factors $S$, $a$ and
$\dd u/\dd x$, not the Lorentzian itself. We use 1601 panels between $10^{-3}$
and $0.85\,R_\odot$; doubling their number changes the flux by less than $0.5\%$
at every mass.}



\bibliography{refs}

\end{document}